%% file: asbl-local.tex
\documentclass{jfm}
\usepackage[utf8x]{inputenc}
\usepackage{color}
\usepackage{graphicx}
\usepackage{amsmath}
\input{header}

\newcommand{\usuck}{V_S}
\newcommand{\e}{ {\mathrm e} }
\renewcommand{\vec}[1]{\mathbf{#1}}

\title[Localized travelling waves in the ASBL]%
{Localized travelling waves in the asymptotic suction boundary layer}
\author[T. Kreilos, J. F. Gibson, T. M. Schneider]%
{Tobias Kreilos$^{a}$\footnote{tobias.kreilos@epfl.ch}, 
John F. Gibson$^{b}$\footnote{john.gibson@unh.edu},
Tobias M.\ Schneider$^{a}$\footnote{tobias.schneider@epfl.ch}}
\affiliation{
$^a$Emergent Complexity in Physical Systems Laboratory (ECPS), 
\'Ecole Polytechnique F\'ed\'erale de Lausanne, 1015 Lausanne, Switzerland \\
$^b$Department of Mathematics and Statistics, University of New Hampshire, 
Durham, New Hampshire 03824, USA}

\date{\today}

\begin{document}

\maketitle

\begin{abstract}
We present two spanwise-localized travelling wave solutions in the asymptotic 
suction boundary layer, obtained by continuation of solutions of plane Couette flow.
One of the solutions has the vortical structures located close
to the wall, similar to spanwise-localized edge states previously found for this system.
The vortical structures of the second solution are located in the free stream far above
the laminar boundary layer and are supported by a secondary shear gradient that is
created by a large-scale low-speed streak. The dynamically relevant eigenmodes of
this solution are concentrated in the free stream, and the departure into turbulence 
from this solution evolves in the free stream towards the walls. 
For invariant solutions in free-stream turbulence, this solution thus shows 
that that the source of energy of the vortical structures can be a dynamical 
structure of the solution itself, instead of the laminar boundary layer.
\end{abstract}


\section{Introduction}

In the last few decades dynamical systems theory has been established as a new paradigm 
for studying turbulence. The foundation for this progress has been the computation of invariant
solutions of the Navier-Stokes equations which capture crucial features of turbulent flows
at moderate Reynolds numbers. Fully 3D and fully nonlinear invariant solutions,
in the form of equilibria, travelling waves, and periodic orbits, have been computed 
for canonical confined shear flows such as pipe flow,
plane Couette 
and plane Poiseuille flow \citep{Kawahara2012}.
Such studies in minimal flow units have produced spatially periodic and thus infinitely 
extended invariant solutions. 
More recently, spatially localized solutions have been 
constructed for flows in large domains \citep{Gibson2014,Brand2014,Zammert2014}.
These localized solutions are better suited to investigate transitional
turbulence, which usually originates in localized turbulent patches
and shows rich spatio-temporal dynamics.

A natural further step is to extend invariant solutions from confined flows to 
external flows and eventually free-stream turbulence. Similarities between external
and internal flows near transition suggest that this extension is possible;
for example, both plane Couette flow and the boundary layer exhibit
roll-streak coherent structures \citep{Nagata1990,Robinson1991}
and localized turbulent spots that grow
into the surrounding laminar flow. 
However, invariant solutions have not yet been computed for developing 
boundary layers. The key difficulty is that the growth of the boundary layer 
thickness with distance from the leading edge breaks the continuous translation 
symmetry of the flow in the downstream direction. This broken symmetry disallows 
downstream travelling wave solutions, which are the dynamically simplest possible invariant 
solutions and the most straightforward to compute.

We will therefore consider a modified, canonical, and well-studied open boundary
layer, the asymptotic suction boundary layer (ASBL). In the ASBL, constant suction 
through a porous wall counteracts the growth of the boundary layer, and far from 
the leading edge a parallel streamwise invariant flow is established \citep{Schlichting2004}.
Several invariant solutions have been computed for the ASBL. Edge-tracking methods 
in a minimal flow unit yield a periodic orbit of very long period \citep{Kreilos2013}. 
In a spanwise extended domain, a spanwise-localized relative periodic orbit has been 
found \citep{Khapko2013,Khapko2014}.
The only known travelling waves of the ASBL have been constructed by \citet{Deguchi2014}
(hereafter referred to as DH14).
Like the periodic orbits of \citet{Kreilos2013}, one of the DH14 
travelling waves is dominated by vortical structures attached to the wall.
The other has vortical structures at a large distance from the wall. DH14 thus 
terms these travelling waves ``wall-mode'' and ``free-stream'' coherent structures, 
respectively. Both of the DH14 ASBL travelling waves are localized in the 
wall-normal direction but periodic in the spanwise and streamwise directions.
The applicability of the high-Reynolds number asymptotics within the framework
of vortex-wave interactions \citep{Hall1991} in growing boundary layers
was investigated in two further papers by the same authors \citep{Deguchi2015,Deguchi2015a}.

In this paper, we demonstrate the existence of wall-mode and free-stream 
travelling-wave solutions of the ASBL that are localized in both the spanwise 
and the wall-normal directions. Like the DH14 solutions, the wall mode has 
tilted, counter-rotating vortices and high- and low-speed streaks that reside
in the near-wall laminar shear region, while the vortices of the free-stream 
lie within the free stream and support streaks closer to the wall.
We show that the free-stream travelling wave is dynamically detached from the wall and supports 
turbulence localized in both the spanwise and the wall-normal direction. While 
transitioning to turbulence, the localized turbulent region in the cross-flow 
plane slowly and almost isotropically invades laminar flow at constant front 
velocity until the wall is reached. 

\section{Methods}
    The ASBL consists of an incompressible fluid streaming
    over a flat plate into which it is homogeneously sucked at a constant rate. Far from
    the leading edge the suction exactly compensates the growth of the boundary
    layer and a translationally invariant laminar profile emerges \citep{Schlichting2004}.
    The laminar downstream velocity increases from zero at the wall to the
    free-stream velocity $U_\infty$, 
    with a characteristic length scale $\delta = \nu/\usuck$ determined by 
    the kinematic viscosity $\nu$ and the suction velocity $\usuck$.
    Denoting the downstream, wall-normal and spanwise directions as $x,y$ and $z$ 
    respectively, and the corresponding components  of velocity as $u,v,w$, the 
    governing Navier-Stokes equations for $\vec u$  and pressure $p$ are
    \begin{equation}
      \partial_t \vec u + (\vec u \cdot \vec \nabla)\vec u
	= -\vec\nabla p + \nu \vec\nabla^2 \vec u,
    \end{equation}
    together with the incompressibility condition
    $\vec \nabla \cdot \vec u = 0$ and the boundary conditions
    \begin{equation}
      \vec u(x,0,z) = (0, -\usuck, 0),\qquad 
      \vec u(x,y\rightarrow\infty,z) = (U_\infty, -\usuck, 0).
    \end{equation}
    The laminar flow profile of the ASBL
    \begin{equation}
     \vec u_L(x,y,z) = (U_\infty ( 1-\e^{-y/\delta} ), -\usuck, 0)
    \end{equation}
    is an analytic solution to this system of equations. 
    The laminar $99\%$ boundary layer thickness, i.e.\ the value of $y$ 
    at which $u(y) = 0.99U_\infty$, is $\delta_{99\%} = 4.605\delta$.
    The Reynolds  number for the ASBL is defined as
    \begin{equation}
      \Rey = \frac{U_\infty \delta}{\nu} = \frac{U_\infty}{\usuck}.
    \end{equation}

    In numerical simulation we enforce the upper boundary condition at a
    finite height $y=H$, leading to a slight modification of the laminar profile,
    \begin{equation}
     \vec u_0(x,y,z) = (U^* ( 1-\e^{-y/\delta} ), -\usuck, 0),
    \end{equation}
    with $U^* =  U_\infty/(1-\e^{-H/\delta})$.
    This profile approaches $\vec u_L$ and the computational
    flow approaches the ASBL as $H\rightarrow\infty$.
    At $H=20\delta$ the value of $\delta_{99\%}$ differs by only $10^{-6}$
    from its value at $H \rightarrow \infty$.
    Our simulations are performed with a parallel version of the
    pseudospectral code channelflow \citep{channelflow}, 
    which simulates the Navier-Stokes equations for incompressible
    fluids in rectangular domains with two periodic 
    directions and no-slip boundary conditions in the third direction. 
    The spatial discretization uses Fourier modes in the streamwise and 
    spanwise directions and Chebyshev polynomials in the wall-normal direction.
    Travelling wave solutions are found with a Newton-Krylov-hookstep algorithm \citep{Viswanath2007,Viswanath2009},
    and predictor-corrector parameter continuation is performed by extrapolating
    three known solutions at different parameter values
    and correcting with Newton's method afterwards.
  
    To construct localized solutions of the ASBL, we perform a
    homotopy continuation from known solutions of plane Couette flow,
    the spanwise-localized equilibria EQ7-1 and EQ7-2 of \citet{Gibson2014}.
    Choosing a coordinate system in which the lower plate at $y=0$ 
    is motionless and the upper plate at $y=H$ moves at speed $U_\infty$,
    a homotopy from plane Couette to ASBL conditions can be defined by 
    increasing $\usuck$ from $0$ to a finite value such that
    $H/\delta \gg 1$.
    Starting at plane Couette conditions with $U_\infty H / \nu = 1600$ and
    $\usuck=0$, we increase $\usuck$, $H$ and $L_x$ and lower $\nu$
    until we arrive at ABSL flow conditions with $\Rey = U_\infty / \usuck \simeq 2000$
    and $H \usuck /\nu = H/\delta = 20$. Then within the ASBL we vary the Reynolds 
    number by increasing $\nu$ and $\usuck$ proportionally so that $\delta = \nu/\usuck$ 
    remains constant.
    In what follows all lengths, velocities, and times are normalized by $\delta, U_\infty,$ 
    and $\delta/U_\infty$.
  
\section{Results}
  \subsection{Spanwise-localized wall-mode and free-stream coherent structures in the ASBL}
    
    We successfully continued both EQ7-1 and EQ7-2 to ASBL conditions with 
    $[L_x, H, L_z]$ at $[4\pi,20,24\pi]$ and $[5.5\pi,20,24\pi]$ respectively, 
    and with a resolution of $[48,121,512]$ grid points.
    In the ASBL these states are spanwise-localized 
    travelling waves similar to the free-stream and wall-mode travelling waves of DH14,
    respectively. We henceforth refer to them as the free-stream coherent 
    structure (FCS) and the wall mode (WM) of the ASBL. After continuation 
    into the ASBL the FCS retains EQ7-1's shift-reflect symmetry
    \begin{equation}
      [u,v,w](x,y,z) = [u,v,-w](x+L_x/2, y, -z)
      \label{eqn:shiftreflect}
    \end{equation}
    and the WM retains EQ7-2's $z$-mirror symmetry
    \begin{equation}
      [u,v,w](x,y,z) = [u,v,-w](x, y, -z).
      \label{eqn:zmirror}
    \end{equation}
    These symmetries were enforced throughout the continuations.

\begin{figure}
      \centering
      \includegraphics{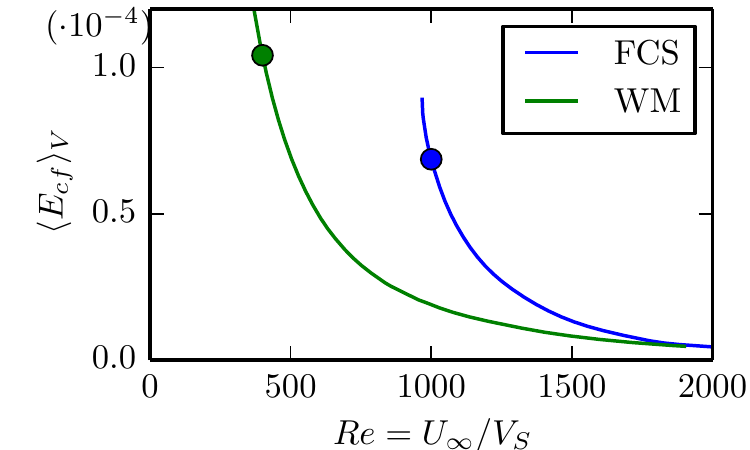}
      \caption{\label{fig:bifdiag}
        Bifurcation diagram of the free-stream (FCS) and wall-mode (WM) solutions 
        showing volume-averaged cross-flow energy $\langle E_{cf}\rangle_{V} = \frac 1V \int_V (v^2 + w^2) \mathrm dx \, \mathrm dy\, \mathrm dz$ 
        as a function of Reynolds number. Both solutions emerge as lower branches in saddle-node 
        bifurcations. The two circles indicate the states depicted in 
        figure~\ref{fig:states_asbl}(c-f). 
      }
    \end{figure}

    Figure~\ref{fig:bifdiag} shows the bifurcation diagram of the FCS and WM
    states in the ASBL with $\Rey$ as control parameter.
    The two states appear as lower branches in saddle-node bifurcations 
    at $\Rey = 968$ for FCS and $\Rey = 348$ for WM. The structure of
    the velocity fields shows little variation with $\Rey$; i.e.
    the states do not change qualitatively along the continuation.  We choose
    to discuss them somewhat above their respective bifurcation points, 
    at $\Rey = 1000$ and $\Rey = 400$, respectively. 
    
    \begin{figure}
      \centering
      \includegraphics[width=.45\linewidth]{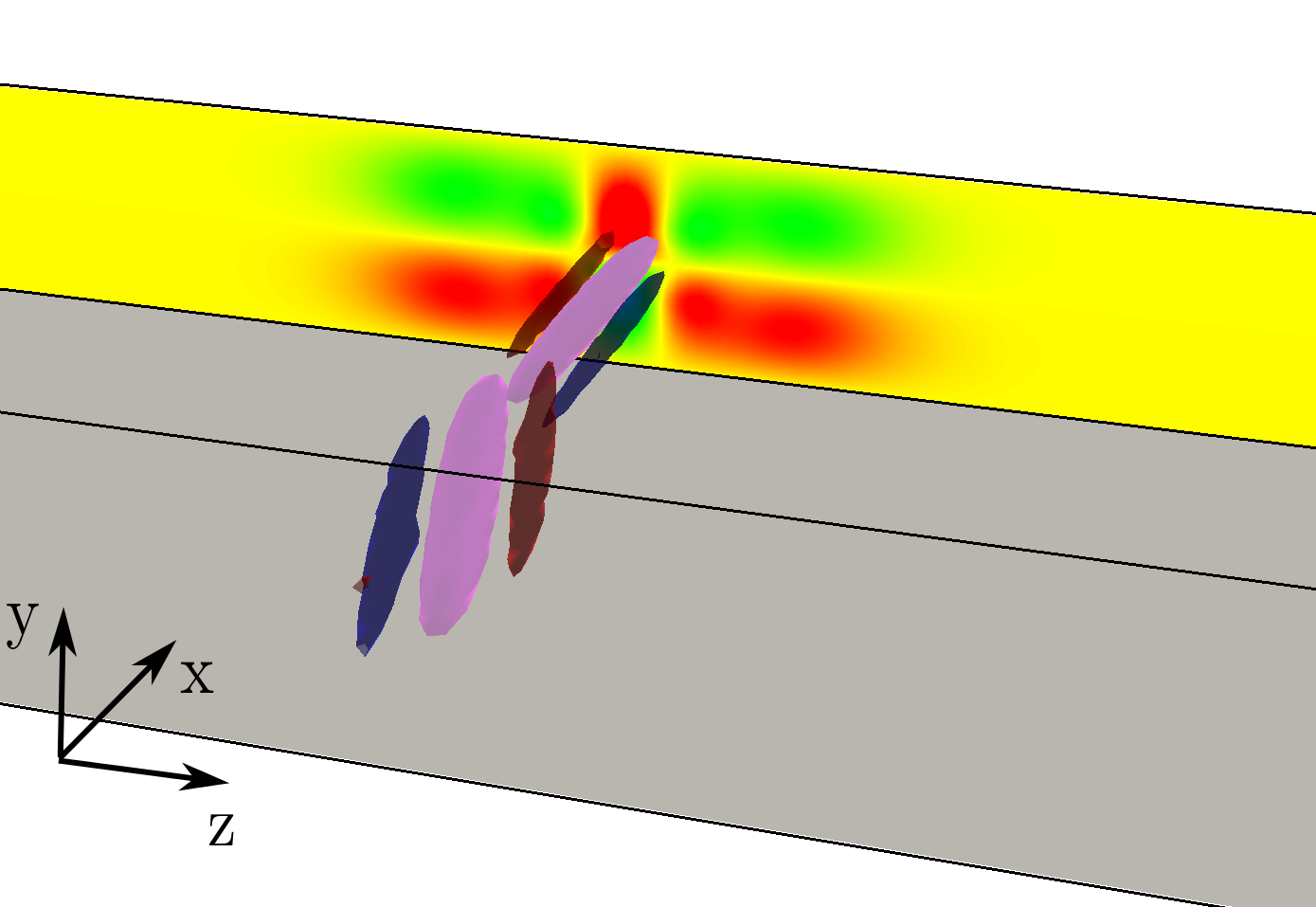}
      \hspace{3mm}
      \includegraphics[width=.45\linewidth]{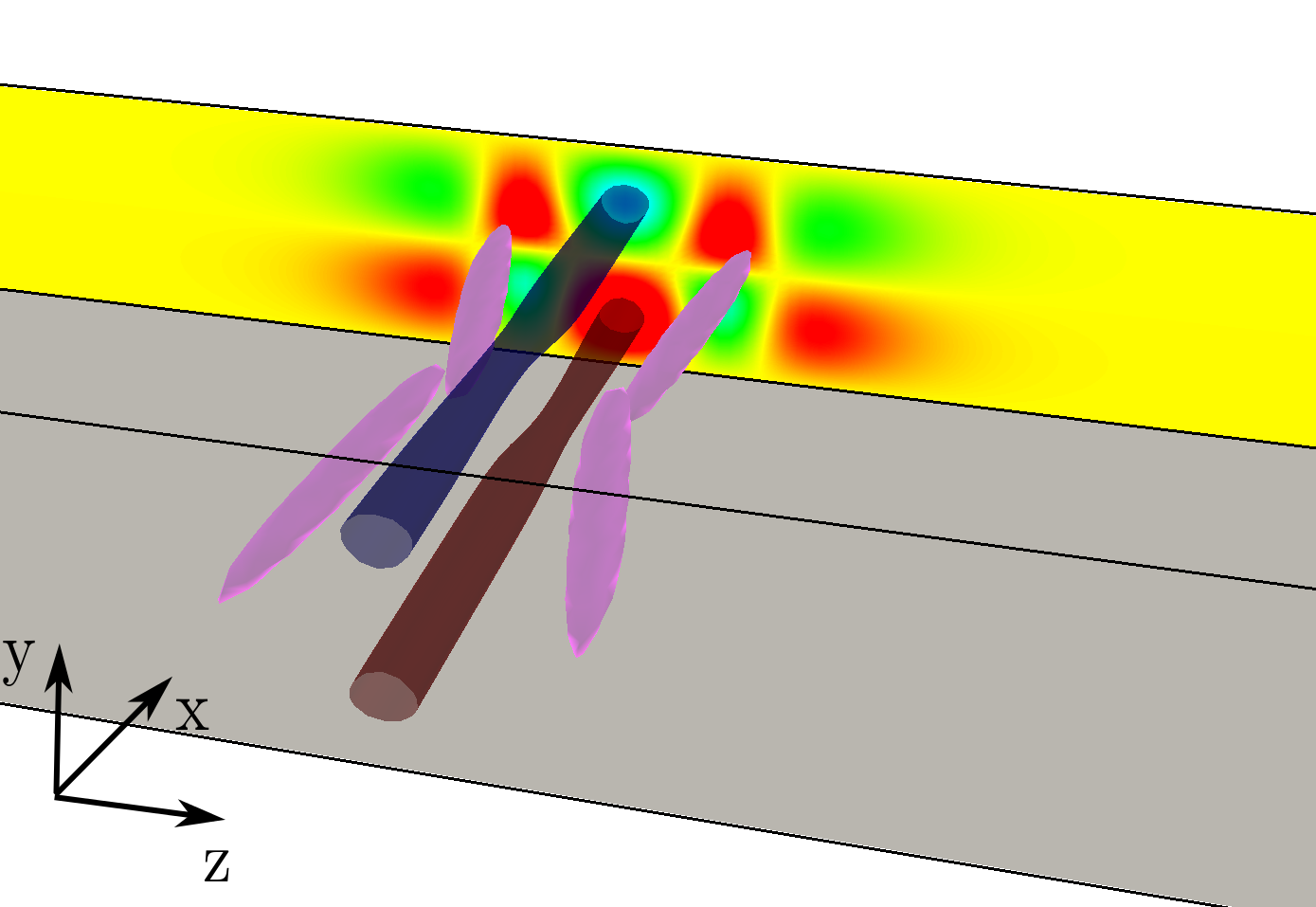}\\
      \includegraphics[width=.45\linewidth]{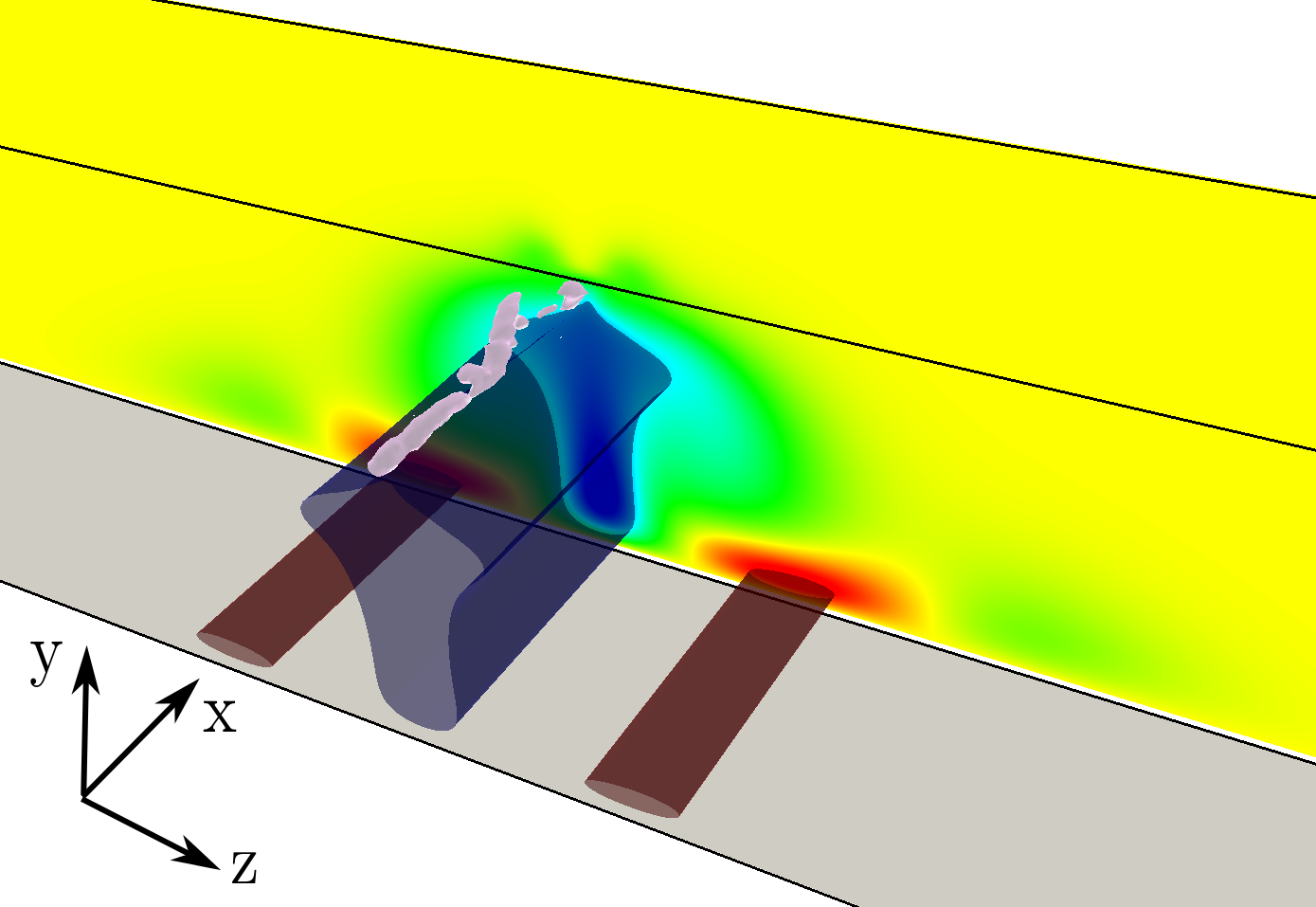}
      \hspace{3mm}
      \includegraphics[width=.45\linewidth]{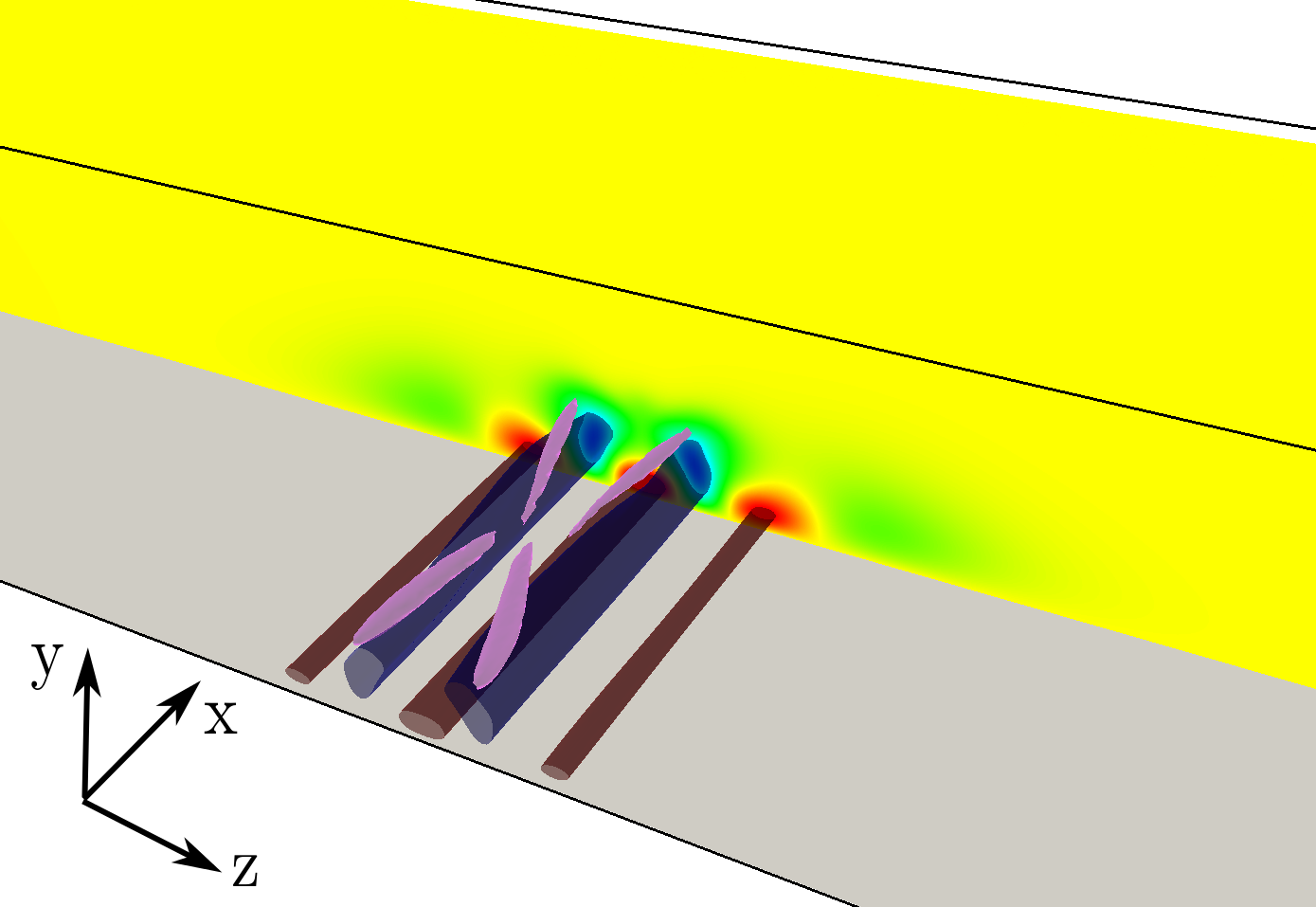}\\
      \includegraphics[width=.45\linewidth]{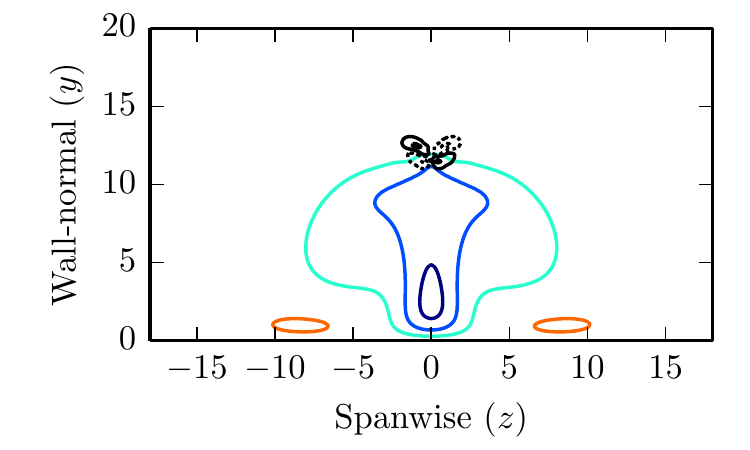}
      \hspace{3mm}
      \includegraphics[width=.45\linewidth]{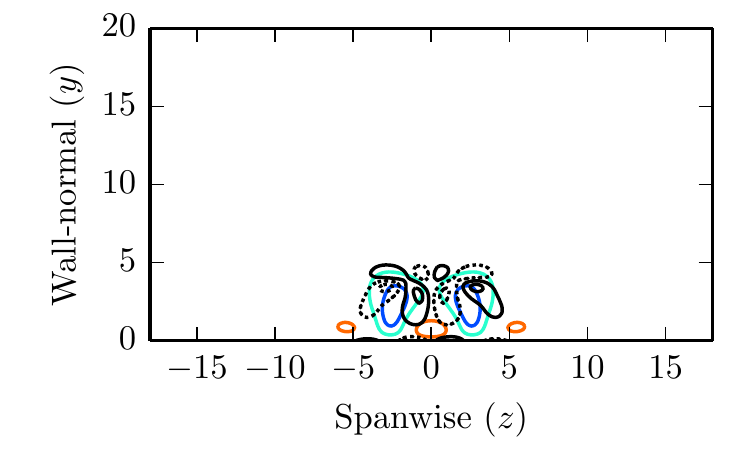}
      \begin{picture}(0,0)
      \put(-375,330){\footnotesize{(a)}}
      \put(-375,220){\footnotesize{(c)}}
      \put(-375,100){\footnotesize{(e)}}
      \put(-188,330){\footnotesize{(b)}}
      \put(-188,220){\footnotesize{(d)}}
      \put(-188,100){\footnotesize{(f)}}
      \end{picture}
      \caption{\label{fig:states_asbl}
        (a,b) Two localized equilibria of plane 
      Couette flow, (a) EQ7-1 and (b) EQ7-2 from \citet{Gibson2014} 
      at $U_{\infty}H/\nu = 1600$. Colours 
      at the boundaries of the box show the streamwise-averaged
      deviation of streamwise velocity from the laminar profile.
      Isocontours of the $\lambda_2$ vortex detection criterion 
      \citep{Jeong1995} are shown in purple. EQ7-1 has one pair of 
      counter-rotating vortices, while EQ7-2 has two pairs.
      High and low-speed streaks are indicated by isocontours of the 
      deviation of streamwise velocity from laminar flow: red for 
      high-speed streaks (faster than laminar) and blue for low-speed 
      streaks. 
      (c,d) Free-stream coherent structure (FCS) and wall-mode (WM) 
      travelling-wave solutions of the ASBL obtained by continuation 
      from EQ7-1 and EQ7-2, with the  same colour coding as in (a,b).
      (e,f) The roll-streak structures of the FCS and WM are
      indicated with isocontours of streamwise-averaged perturbation 
      velocity $u$ (colours) and streamwise-averaged vorticity $\omega_x$ 
      (solid black for positive $\omega_x$, dotted for negative).
      The FCS shown in (c,e) at $Re=1000$ consists of two weak high-speed 
      streaks close to the wall and one low-speed streak that extends 
      far into the free stream. The vortices are localized at 
      $y \simeq 12$, far above the laminar boundary layer.
      The WM shown in (d,f) at $Re=400$ contains two pairs of vortices, 
      two low- and three-high speed streaks, all within the 
      laminar boundary layer, $y \leq \delta_{99\%} \simeq 5$, 
      }
    \end{figure}
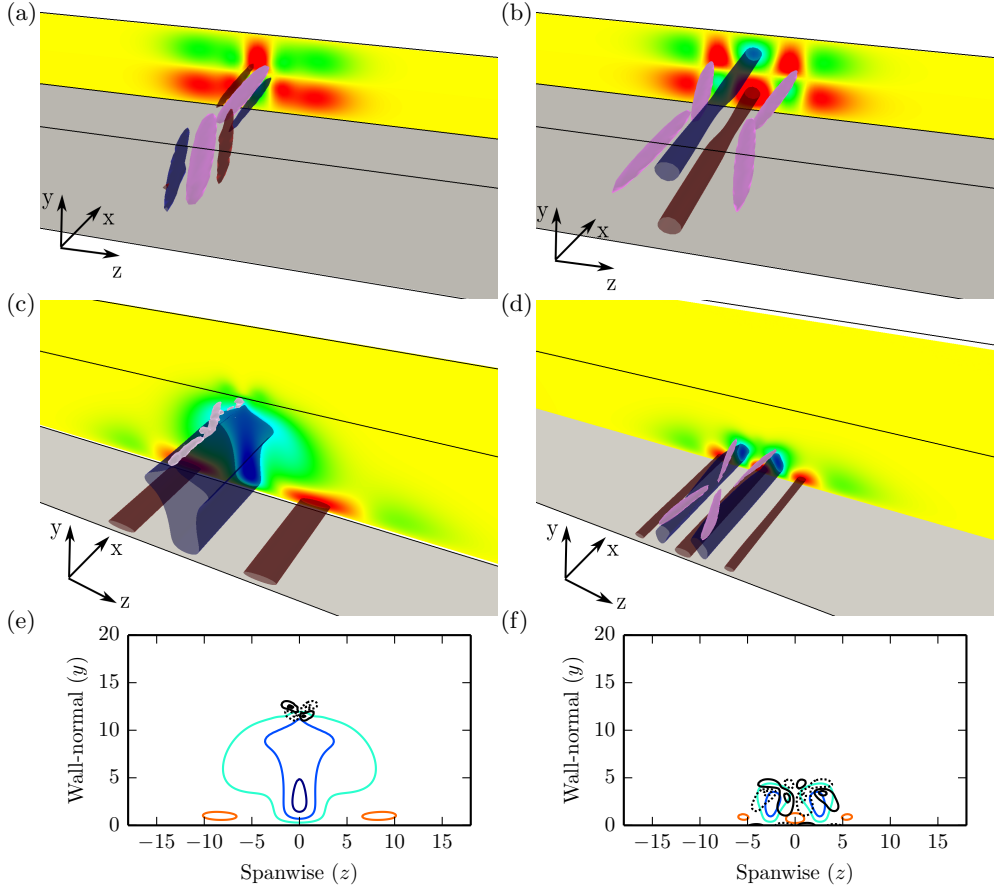
    
\begin{figure}
      \centering
      \includegraphics[width=.48\linewidth]{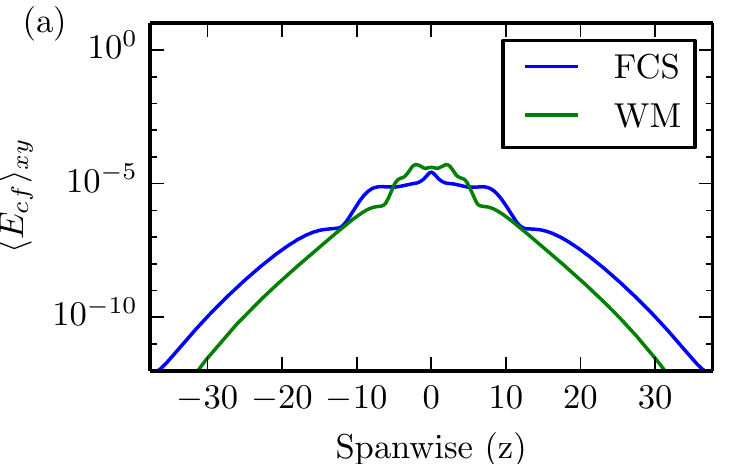}
      \includegraphics[width=.48\linewidth]{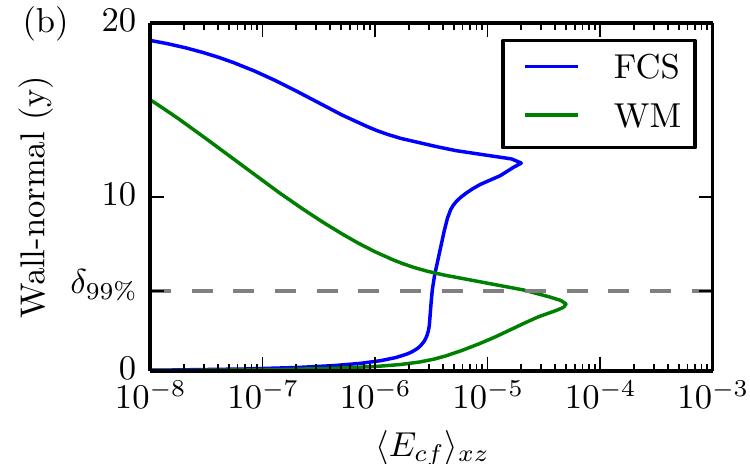}
      \caption{\label{fig:localization}
        Localization properties of the FCS (blue) and WM (green) in the (a) spanwise
        and (b) wall-normal directions. Localization is indicated by the cross-flow 
        energy  $E_{cf} = v^2 + w^2$ averaged in $x,y$ and $x,z$, respectively.
        The exponential drop-off in cross-flow energy with (a) $z$ and (b) $y$
        confirms that the computational domain is sufficiently large.
        (b) The peak of cross-flow energy for the FCS is at $y \simeq 12$, far above
        the laminar boundary layer $y \leq \delta_{99\%} \simeq 5$, while for WM 
        the peak is within it.
      }
    \end{figure}

    Figure~\ref{fig:states_asbl} compares the ASBL states to their 
    progenitor states in plane Couette flow.
    Figure~\ref{fig:states_asbl}(a,b) shows EQ7-1 and EQ7-2 of plane Couette flow
    and (c,d) the corresponding FCS and WM states obtained by continuation 
    to the ASBL,
    with vortices indicated by isosurfaces of $\lambda_2$ \citep{Jeong1995} in purple 
    and velocity fluctuations by isosurfaces of streamwise velocity in blue and red 
    (see figure caption for details).
    Figure~\ref{fig:states_asbl}(e,f) shows the roll-streak-structure of FCS and
    WM in the ASBL by isocontours of the downstream velocity (colours) and downstream 
    vorticity (black).
    The first important observation is that both states are well-localized
    in the spanwise direction to within a range of $|z| < 10$, considerably 
    smaller than boundaries of the computational domain at $L_z/2 = 12\pi \simeq 38$ 
    (see also figure~\ref{fig:localization}).
      
    Like EQ7-1 in plane Couette flow, the FCS in the ASBL consists of a
    single pair of staggered and tilted counter-rotating vortices. These vortices are located
    at $y\simeq12$, far above the laminar boundary layer thickness at $\delta =1$
    or the laminar $99\%$ boundary layer thickness at $\delta_{99\%} = 4.6$. 
    Located below the vortices is a large low-speed streak, i.e. a region
    where the streamwise velocity is slower than in the surrounding laminar flow.
    The low-speed streak originates at the wall and ends below the vortices at $y \approx 12$,
    creating a small shear from which the vortices draw energy. 
    Note that the streak contour lines are nearly as dense near the vortices as 
    at the wall, indicating that the shear near the vortices is large, and that the large 
    wall-normal extent of the low-speed streak in figure~\ref{fig:states_asbl}(e) is not 
    an artifact of the choice of contour levels. Thus a key property of the FCS is
    that its vortices are sustained not by the laminar shear near the wall, but by
    shear far from the wall that the solution creates itself. The vortices push fluid to the 
    sides by linear advection, creating a weak but large-scale circulation. 
    This circulation pushes fluid from the free stream towards the wall on the
    right and left sides of the vortices, creating two weak high-speed streaks, and pulls 
    fluid up in the center, creating the low speed-streak.
    The speed of the travelling wave is $c_{FCS} = 0.905$, notably fast compared 
    to the speed of turbulent spots in boundary layers (which travel at roughly 
    $c=\frac 23$, \citet{Levin2007}) but consistent with the $1-c \sim 100 \cdot O(\Rey^{-1})$
    scaling estimated by DH14. Note that the laminar profile reaches this wavespeed 
    ($u_L(y) = c_{FCS}$) at $y = 2.35$.

    There are several notable similarities between the spanwise-localized 
    FCS figure~\ref{fig:states_asbl} and the spanwise-periodic free-stream coherent 
    structure of DH14. Both are derived from the EQ7 plane Couette solution of 
    \citet{Gibson2009}. The shift-reflect symmetry (\ref{eqn:shiftreflect}) of the 
    FCS is the symmetry that remains after breaking the $z$-periodicity of the 
    DH14 free-stream solution (their equation 3.4). Comparison of 
    figure~\ref{fig:states_asbl}(e) with figure 3(e) of DH14 reveals similar 
    structure and wall-normal position of the counter-rotating vortices
    of the two solutions, despite the differences in streamwise wavelength 
    and Reynolds number ($L_x = 4\pi, \Rey=968$ here versus $L_x=10\pi, \Rey=80,000$ 
    in DH14). In both solutions the slow-speed streaks extend farther into the free 
    stream than do the high-speed streaks, although for the DH14 free-stream solution,
    there is a clear concentration near the wall and decay away from it, whereas the 
    low-speed streaks of the FCS have a more uniform core that extends farther into 
    the free stream.

    The WM in the ASBL is notably different from the FCS. The WM has 
    two pairs of staggered, counter-rotating vortices much closer to the wall, 
    below $y=\delta_{99\%} = 4.65$.
    By linear advection (lift-up effect), the vortices create and sustain 
    two low-speed and three high-speed streaks which are also very near the wall.
    The vortices are slightly tilted and inclined in the downstream direction,
    a pattern that is commonly observed in wall-bounded shear flows \citep{Adrian2007}.
    With all structures closely attached to the wall and the pattern of two
    vortices leaning over a low-speed streak, the WM is similar to the
    spanwise-localized edge states of \citet{Khapko2013}, although in 
    the latter, the crossing of the vortices over the streak destabilizes the 
    streak, leading to breakup and reformation at a different position, 
    and resulting in a time-dependent, relative periodic orbit solution. 
    The speed of the WM  travelling wave is $c_{WM} = 0.87$, a speed
    reached by the laminar profile at $y=2.04$. The slower wavespeed 
    of the WM compared to the FCS is consistent with its closer proximity to 
    the wall. The WM is clearly related to the TW2-2 near-wall travelling wave 
    of plane Poiseuille flow of \citet{Gibson2014}, both in their origin from 
    EQ7-2 and in the structure of their vortices and streaks. How the WM relates 
    to the wall mode of DH14 is less clear. The latter was continued from a 
    different solution of sliding Couette flow and has vorticity concentrated
    much closer to the wall (comparing figure~\ref{fig:states_asbl}(d,e) to 
    DH14 figure 3(b), albeit at different parameter values).

    At a broad level, the FCS and the WM are composed of the same typical 
    flow structures -- streamwise streaks and downstream-oriented vortices -- as 
    almost all known invariant solutions in shear flows. 
    The WM displays the familiar interactions known from the
    self-sustaining process \citep{Hamilton1995,Waleffe1997,Schoppa2002}
    or the vortex-wave interaction of \citet{Hall2010}.
    In the FCS, however, the strong vortices are not supported by a linear
    instability of the underlying streak (DH14), as complex
    nonlinear interactions dominate.
    
  \subsection{Dynamical properties of the free-stream coherent structure}
\begin{figure}
      \centering
      \includegraphics[width=.6\linewidth]{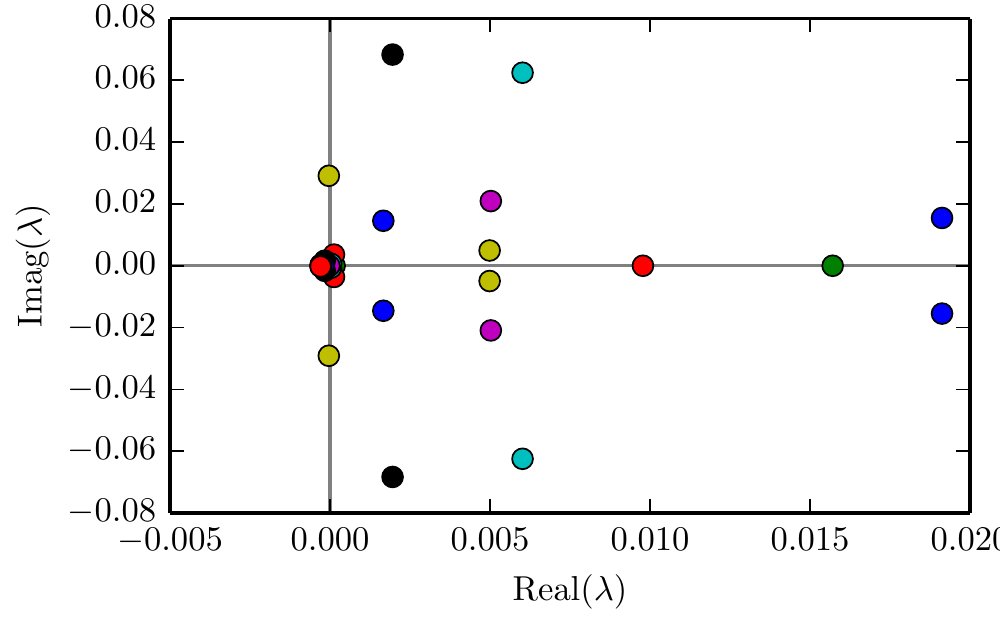}
      \caption{\label{fig:FCS:spectrum}
       The leading eigenvalues of the FCS at $Re=1000$ with 
       restriction to the shift-reflect symmetry subspace (\ref{eqn:shiftreflect}).
       With an increased resolution of $[64,121,576]$ grid points, the leading
	10 eigenvalues differ by $10^{-4}$, confirming that the solution is well converged.
      }
    \end{figure}

\begin{figure}
      \centering
      \includegraphics[width=.48\linewidth]{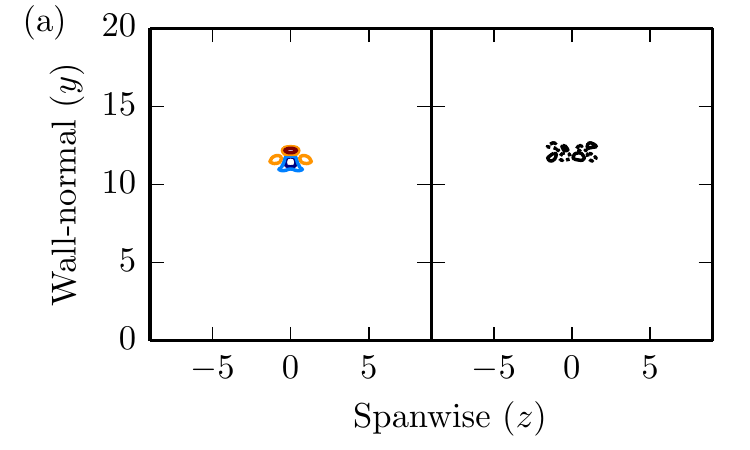}
      \includegraphics[width=.48\linewidth]{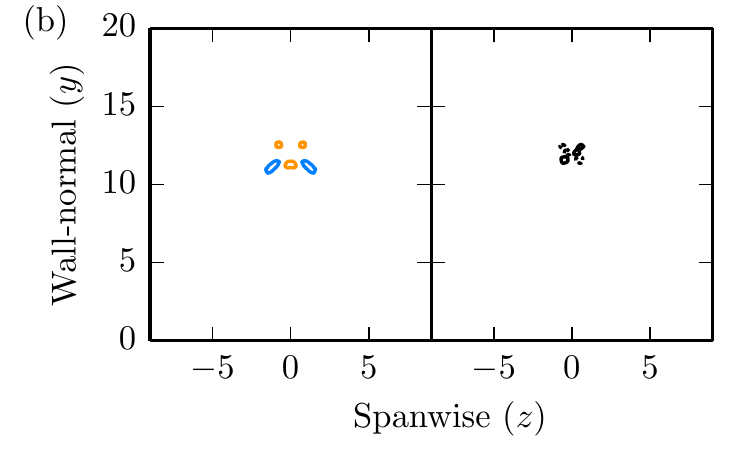}
      \includegraphics[width=.48\linewidth]{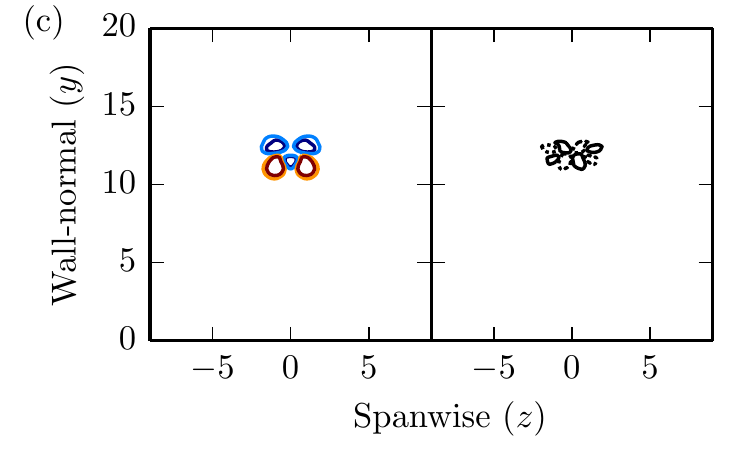}
      \includegraphics[width=.48\linewidth]{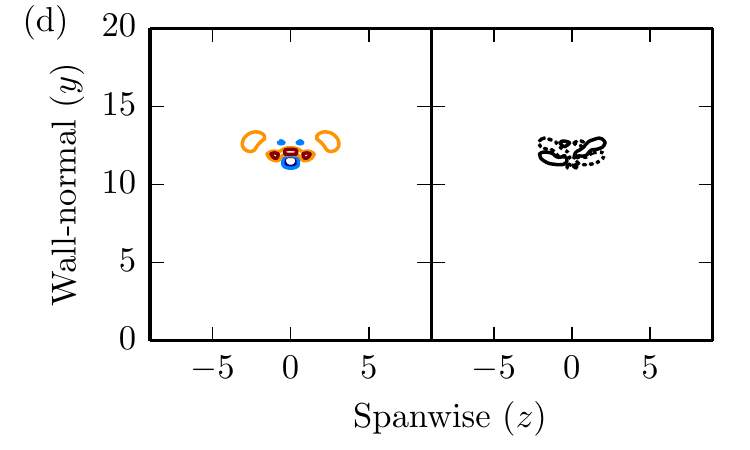}
      \caption{\label{fig:FCS:eigenmodes}
        The four leading eigenmodes of the FCS. For complex eigenmodes, 
        the real part is shown. The left part of each plot 
        shows isocontours of the streamwise-averaged deviation of streamwise velocity 
        from laminar flow, the right part the isocontours of streamwise-averaged
        streamwise vorticity. For all eigenmodes, the streaks and roll structures 
        are localized in the free-stream, close to the vortices of the FCS.
      }
    \end{figure}

    In this section we show that the dynamical evolution of the FCS
    is detached from the wall. Figure~\ref{fig:FCS:spectrum} shows
    that the FCS has $O(10)$ unstable modes (eigenvalues with positive 
    real part). The edge states of \citet{Khapko2013,Khapko2014} in comparison 
    have a single unstable eigenvalue by construction.
    Figure~\ref{fig:FCS:eigenmodes} shows the roll-streak structure of the 
    four leading unstable eigenmodes. These are the most dynamically 
    relevant eigenmodes, since they will dominate the evolution of trajectories
    that approach the FCS transiently. The leading eigenmodes are 
    concentrated around the vortical structures of FCS at 
    $y\simeq12$, far from the wall. The linear dynamics near the
    FCS is thus concentrated near the free-stream part of the
    solution, and the streaks that extend close to the wall do not play
    a significant role.
    
    \begin{figure}
      \centering
      \includegraphics[width=.48\linewidth]{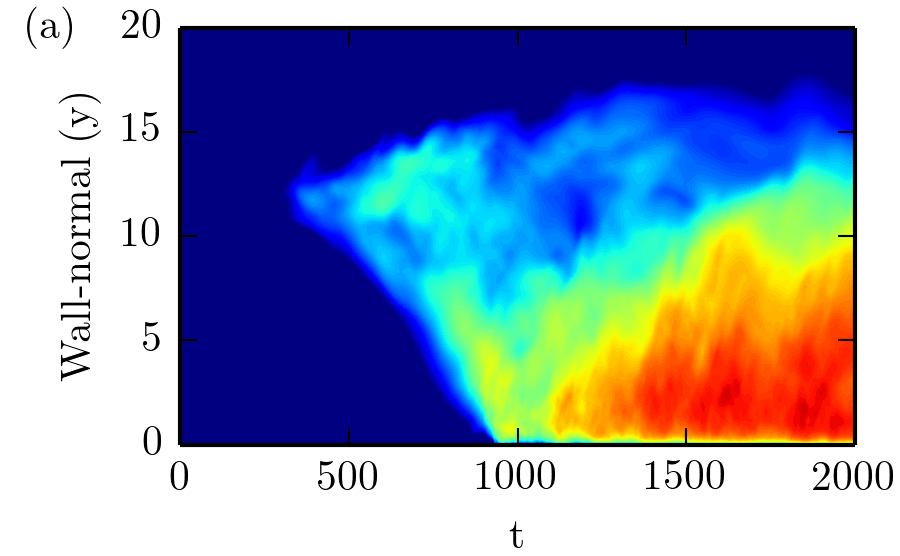}
      \includegraphics[width=.48\linewidth]{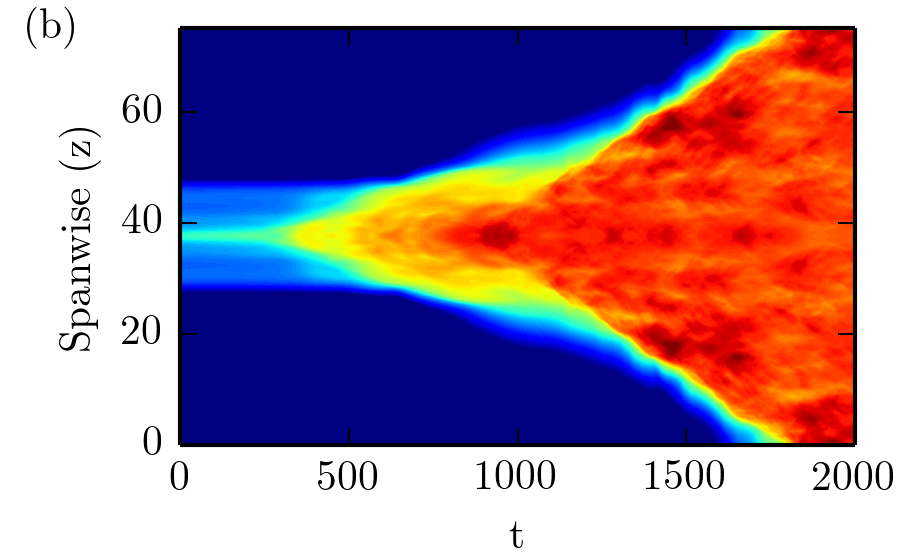}
      \caption{\label{fig:FCS:evolution}
        Space-time plots of the transition from FCS to turbulence at $Re=1000$.
        The averaged cross-flow energy $E_{cf}$ for the growth of a small perturbation
        of the FCS is shown as a function of time and the (a) wall-normal and (b)
        spanwise directions. The perturbation remains small for about $350$ advective 
        time units. After $t\simeq350$ the perturbations grow slowly 
        in the spanwise direction and from the free stream towards the wall. 
        When perturbations reach the laminar boundary layer $y\leq \delta_{99\%}$
        around $t \simeq 800$, they strengthen and later spread rapidly along the wall, 
        as shown by the increased slope of the red area in (b).
      }
    \end{figure}
    
    To study dynamics of transients near the FCS, we perturb the FCS slightly and 
    follow the time evolution of the perturbed initial condition. Here we present
    results obtained by rescaling FCS by a factor of $1.01$; the qualitative 
    results are unchanged if a small random perturbation is added instead.
    A movie in the supplementary material shows the evolution in time.
    Figure~\ref{fig:FCS:evolution} shows space-time plots of the cross-flow
    energy $E_{cf} =  v^2 + w^2$ \citep{Kreilos2013}
    as a function of $y$ and as a function of $z$, in each case
    averaged over the remaining two directions.
    No changes are visible for the first 350 time units as the exponential
    amplifications of the linearly unstable modes grow to large amplitudes.
    (This time of course depends on the magnitude of the initial perturbation).
    At $t\simeq 350$ velocity perturbations near the vortical structures 
    at $y\simeq12$ and $z=0$ grow to amplitudes comparable to the FCS 
    velocity field. These perturbations spread in the $y$ direction mostly 
    towards the wall and in the spanwise direction at roughly equal 
    but slow propogation speeds. This slow spreading continues
    until the fluctuations reach the laminar boundary layer at 
    $t \simeq 900$. After this, the fluctations strengthen and 
    energetic turbulence begins to spread along the wall and also back into 
    free stream. The rate of spanwise spreading is notably faster after the 
    perturbations for $t \geq 1000$.
    This indicates that there are two different mechanisms by which turbulence
    spreads from the FCS: a slow growth of weak perturbations in the free 
    stream followed by a faster, more energetic growth fueled by the shear of
    the laminar boundary layer.

    \begin{figure}
      \centering
      \includegraphics[width=.98\linewidth]{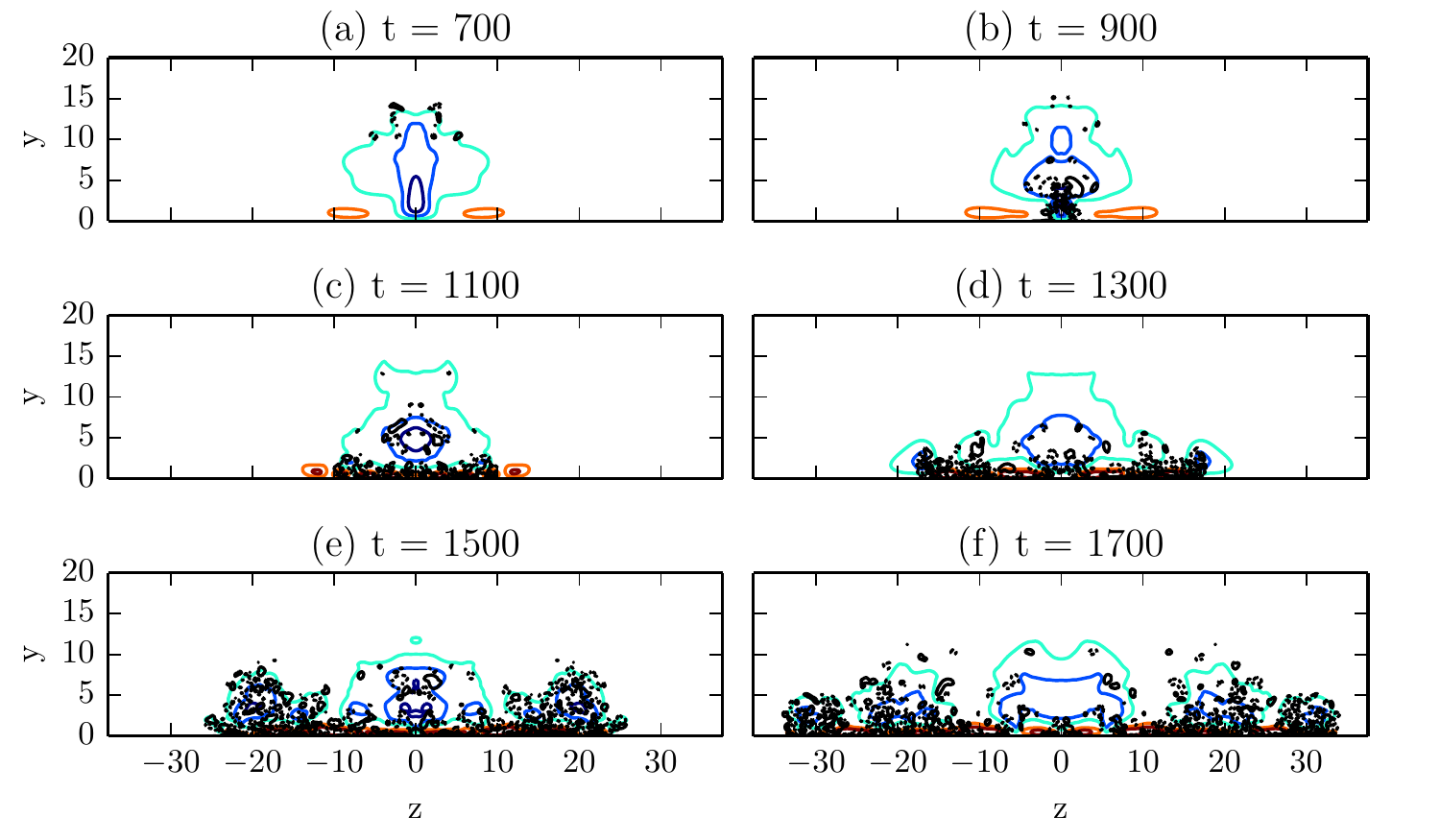}
      \caption{\label{fig:FCS:evolution_contours}
        Time evolution of FCS, illustrating the spreading of weak perturbations 
        towards the wall, followed by growth in energy and faster spreading in the
        spanwise direction. The plots show isocontours of the downstream velocity 
        fluctuations in colours and isocontours of the downstream
        vorticity in black. 
      }
    \end{figure}
    
    The observation that turbulence spreads rapidly once it reaches the wall 
    is confirmed by the plots of the roll-streak structure
    in figure~\ref{fig:FCS:evolution_contours}, with six snapshots of the 
    trajectory taken at equidistant time intervals $\Delta t = 200$.
    In panel (a) the state has not changed much from FCS 
    and the vortical structures
    (black isocontours) are located high in the free stream. In panel (b)
    the perturbations have grown to the laminar boundary layer,
    where they are strongest. In panels (c-f) the stronger perturbations 
    spread in the spanwise direction and start to grow
    in the wall-normal direction, creating a larger turbulent boundary layer, 
    in which disturbances would ultimately grow to reach the upper
    boundary of the computational domain \citep{Schlatter2011}.

\section{Conclusions}
 
  We have presented two travelling wave solutions in the asymptotic suction
  boundary layer which are localized in both the wall-normal and 
  the spanwise directions.
  This is the first time that spanwise-localized invariant solutions
  have been computed in a boundary layer. The wall-mode solution is 
  closely attached to the wall, while the free-stream solution has vortices 
  far from the wall, within the free stream. The vortices of the free-stream 
  solution cannot draw energy from the laminar boundary layer; instead they 
  are supported by a secondary shear gradient created by a large-scale, low-speed 
  streak which stretches from the wall far into the free stream, convecting energy
  from the laminar gradient into the free-stream.
  The identification of this solution is thus an important step for invariant solutions 
  in free-stream  turbulence, as it shows the source of the shear gradient is not 
  necessarily the background laminar flow profile, but can be a dynamical structure of the solution itself.

  The localization of the free-stream solution's unstable eigenfunctions
  and the dynamical evolution of perturbations from it show that the 
  active region of this state is also detached from the wall. 
  Perturbations of the free stream solution grow first around the vortical 
  structures and then spread almost isotropically in the cross-stream plane, 
  similar to a turbulent spot in a wall-parallel plane. Once the strong 
  laminar shear at the wall is reached, the turbulent fluctuations sharply 
  increase in magnitude and spread rapidly along the wall. 

\begin{acknowledgments}
  This work was supported by the Swiss National Science Foundation under grant 
  no.\ 200021-160088. 
\end{acknowledgments}


\input{asbl-local.bbl}
\end{document}

%% file: header.tex
\usepackage{graphicx}
\usepackage{natbib}

\ifCUPmtlplainloaded \else
  \checkfont{eurm10}
  \iffontfound
    \IfFileExists{upmath.sty}
      {\typeout{^^JFound AMS Euler Roman fonts on the system,
                   using the 'upmath' package.^^J}%
       \usepackage{upmath}}
      {\typeout{^^JFound AMS Euler Roman fonts on the system, but you
                   dont seem to have the}%
       \typeout{'upmath' package installed. JFM.cls can take advantage
                 of these fonts,^^Jif you use 'upmath' package.^^J}%
      }
  \else
  \fi
\fi

\ifCUPmtlplainloaded \else
  \checkfont{msam10}
  \iffontfound
    \IfFileExists{amssymb.sty}
      {\typeout{^^JFound AMS Symbol fonts on the system, using the
                'amssymb' package.^^J}%
       \usepackage{amssymb}%
         \let\leq=\leqslant
         \let\geq=\geqslant
      }{}
  \fi
\fi

\ifCUPmtlplainloaded \else
  \IfFileExists{amsbsy.sty}
    {\typeout{^^JFound the 'amsbsy' package on the system, using it.^^J}%
     \usepackage{amsbsy}}
    {}
\fi

\newcommand\Rey{\mbox{\textit{Re}}}  

\newsavebox{\astrutbox}
\sbox{\astrutbox}{\rule[-5pt]{0pt}{20pt}}